\documentclass[aps,prb,twocolumn,groupedaddress,showpacs]{revtex4}
\usepackage{graphicx,amsmath}
\usepackage{amssymb}

\begin{document}
\title{Competition between disorder and Coulomb interaction in
a two-dimensional plaquette Hubbard model}

\author{Hunpyo Lee}
\affiliation{School of General studies, Kangwon National University, Samcheok-Si, 245-711,
South Korea}
\author{Harald O. Jeschke}
\author{Roser Valent\'\i}
\affiliation{Institut f\"ur Theoretische Physik, Goethe-Universit\"at
Frankfurt, Max-von-Laue-Stra{\ss}e 1, 60438 Frankfurt am Main, Germany}
\date{\today}

\begin{abstract}
  We have studied a disordered $N_{\rm c} \times N_{\rm c}$ plaquette
  Hubbard model on a two-dimensional square lattice at half-filling
  using a coherent potential approximation (CPA) in combination with a
  single-site dynamical mean field theory (DMFT) approach with a
  paramagnetic bath. Such a model conveniently interpolates between
  the ionic Hubbard model at $N_{\rm c}=\sqrt{2}$ and the Anderson
  model at $N_{\rm c} = \infty$ and enables the analysis of the
  various limiting properties.  We confirmed that within the CPA
  approach a band insulator behavior appears for non-interacting
  strongly disordered systems with a small plaquette size $N_{\rm c} =
  4$, while the paramagnetic Anderson insulator with nearly gapless
  density of states is present for large plaquette sizes $N_{\rm
    c}=48$.  When the interaction $U$ is turned on in the strongly
  fluctuating random potential regions, the electrons on the low
  energy states push each other into high energy states in DMFT in a
  paramagnetic bath and correlated metallic states with a
  quasiparticle peak and Hubbard bands emerge, though a larger
critical interaction $U$ is needed to obtain this state from the paramagnetic Anderson
  insulator ($N_{\rm c}=48$) than from the band insulator ($N_{\rm
    c}=4$). Finally, we observe a Mott insulator behavior in the
  strong interaction $U$ regions for both $N_{\rm c}=4$ and $N_{\rm
    c}=48$ independent of the disorder strength. We discuss the application
of this model to real materials.
\end{abstract}

\maketitle

\section{Introduction\label{Introduction}}
The subtle interplay among kinetic energy, electronic correlation, a
periodic ionic potential, and disorder in a two-dimensional electronic
system has been an important research topic for several decades and is
still under debate~\cite{Imada1998,Evers2008}. The metal-insulator
transition driven by only one energy scale in half-filled
two-dimensional systems seems to be relatively well understood: (i)
The on-site Coulomb interaction opens a Mott gap, (ii) the periodic
ionic potential opens a band gap and (iii) disorder induces an
Anderson gap~\cite{Anderson1958,Abrahams1979,Kramer1993,
  Slevin1999,Ekuma2014,Ekuma2014b}.  However, this is not the case when
interactions and disorder compete with electronic itinerancy. There
has been a lot of progress on analytical and numerical approaches
employed to tackle theoretically the complexity of competing
interactions and disorder~\cite{Dobrosavljevic1997,Byczuk2005,
  Byczuk2009, Heidarian2004,
  Shinaoka2009,Chiesa2008,Oliveira2014,Ekuma2015, Dobrosavljevic2003}
but a full understanding of this intricate problem is difficult to
achieve.  Moreover, the existence of a rich variety of real systems
where correlation and disorder play an important role such as
disordered perovskite compounds, layered dichalcogenide 1T-TaS$_2$
with Cu intercalation, Sr$_2$Ir$_{1-x}$Rh$_x$O$_4$ at low doping, 
nano-arrays on two-dimensional surfaces~\cite{Kim2005,
  Maiti2007,Lahoud2014,Chikara2015}, granular deposits of transition
metal-based systems~\cite{Muthu2012,Muthu2012_2},
or two-dimensional metal-oxide-semiconductor field-effect
transistors~\cite{Kravchenko1996,Lin2015} to mention a few, calls for
further analysis of this problem.

The ionic Hubbard model with a periodic on-site potential has been
intensively studied via various numerical methods such as quantum
Monte Carlo and cluster-dynamical mean field theory
approaches~\cite{Kancharla2007,Paris2007,Go2011,Tugushev1996,Caprara2000}.  The results of
these studies differ mostly in details.  Overall they describe
metallic, bond order, band insulating, antiferromagnetic and
paramagnetic (PM) Mott insulating phases. On the other hand, in the
absence of Coulomb interactions, the presence of a random disorder
that breaks the periodicity of the on-site ionic potential induces a
different phenomenon; if the hopping strength of the electrons on a
three-dimensional non-interacting system is larger than the
fluctuations induced by the random disorder, an Anderson insulator to
metal transition appears, while the electrons are always confined in a
random potential in one- and two-dimensional non-interacting systems,
regardless of the disorder strength~\cite{Abrahams1979}.  More
controversial is, however, the behavior of the disordered system in
the presence of Coulomb interactions.

Here we investigate some aspects of this problem by considering a
disordered two-dimensional $N_{\rm c} \times N_{\rm c}$ plaquette
Hubbard model on the square lattice at half-filling.  Note that by
considering effects of interactions and different plaquette
sizes this study goes beyond
non-interacting Anderson model and ionic Hubbard model studies.  The
ionic and Anderson models in the non-interacting limit are recovered
for $N_{\rm c}=\sqrt{2}$ and $N_{\rm c} = \infty$, respectively. The
ionic model with $N_{\rm c}=\sqrt{2}$ has a gapped density of states
$\rho(\omega)$ while $\rho(\omega)$ is gapless and shows a flat
behavior around the Fermi level in a coherent potential approximation
(CPA) for $N_{\rm c} = 48$.  These features are identified as an
Anderson insulator behavior.  We have systematically studied in the
framework of single-site dynamical mean field theory (DMFT) with a
paramagnetic (PM) bath~\cite{Georges1996} how $\rho(\omega)$ and the
quasiparticle weight $Z$ evolves either from a PM metal, Anderson
insulator, or a band insulator to a Mott insulator via tuning of the
plaquette size $N_{\rm c}$, the Coulomb interaction $U$ and the
disorder strength $\Delta$.

We will show that (i) at moderate values of $U$ the system presents a
first-order PM metal to Mott insulator transition in the small
randomness regime with $\Delta/t = 1$ and $2$ for both plaquette sizes
$N_{\rm c}=4$ and $48$, in qualitative agreement with former DMFT
results for the system without disorder~\cite{Park2008}.  (ii) The
non-interacting system with $N_{\rm c}=4$ and $48$ in the strongly
fluctuating potential regions behaves as a band insulator and an
Anderson insulator, respectively, and the electrons occupy the low
energy states.  When interactions are turned on, we find that the
electrons lying in low energy states push each other into high energy
states and the system becomes a correlated metal with a quasiparticle
peak and Hubbard bands in the moderate interaction $U$ region.  The
critical interaction $U_{crit}$ where the insulator to correlated metal
transition occurs is larger in the $N_{\rm c}=48$ than in the $N_{\rm
  c}=4$ system. (iii) Finally, in the large interaction region, the
systems are Mott insulators independent of disorder strength and
plaquette size.

\section{Model\label{Model}}
We consider the following Hamiltonian:
\begin{equation}
H = -t \sum_{<i,j>,\sigma} (c_{i\sigma}^{\dagger} c_{j\sigma} + \text{H.c.}) -
\sum_{i,\sigma} (\mu - \epsilon_i) n_{i\sigma} + U \sum_i
n_{i\uparrow}n_{i\downarrow},
\end{equation}
where $t$ describes the electron hopping strength between nearest
neighbors, $\epsilon_i$ is the on-site energy driven by the random
potential at site $i$, $U$ is the repulsive Coulomb interaction, $\mu$
is the chemical potential which is given as $\mu = \frac{U}{2}$ at
half-filling, and $c_{i\sigma}^{\dagger}$ and $c_{i\sigma}$ are the
electron creation and annihilation operators at site $i$ with spin
$\sigma$, respectively. The on-site energies $\epsilon_i$ on the
$N_{\rm c} \times N_{\rm c}$ plaquette sites are sampled randomly from
the interval $[-\Delta / t : \Delta / t]$, where $\Delta$ is the
strength of disorder.  The translational invariance of a $N_{\rm c}
\times N_{\rm c}$ plaquette for a given random disorder is conserved
in the system. We set the hopping strength $t=1$ and the temperature
to $T/t=0.025$ for all single-site dynamical mean field theory (DMFT)
calculations with a paramagnetic bath~\cite{Georges1996}.  We employ a
continuous-time quantum Monte Carlo algorithm as the DMFT impurity
solver~\cite{Rubtsov2005, Gull2011} in our own
implementation~\cite{Lee2014}.  Disorder effects are evaluated with
the coherent potential approximation~\cite{Ekuma2015, Soven1967}.

\section{Results\label{Results}}

\subsection{Non-interacting disordered systems}

While a band insulator appears in the non-interacting ionic model
limit with $N_{\rm c} = \sqrt{2}$ at finite $\Delta / t$, an Anderson
insulator is observed in a two-dimensional non-interacting disordered
system when $N_{\rm c}=\infty$ at finite $\Delta / t$, according to a
standard scaling theory of localization~\cite{Abrahams1979}.  On the
other hand, since we are considering finite $N_{\rm c} \times N_{\rm
  c}$ plaquettes, we estimate that the finite size of the plaquette
may disturb a localization of electrons in the two-dimensional
non-interacting system with weak disorder whenever the Anderson
localization length exceeds the size of plaquette.  In order to
investigate this behavior, we have studied the non-interacting
two-dimensional case both, in the weak and strong disordered regions
as a function of plaquette size $N_{\rm c}$ with the CPA method.
  
While it has been shown that algebraic averaging one-particle
quantities within the CPA approach fail to account for the Anderson
localization in the disordered system exactly unlike the recently
developed cluster typical medium theory\cite{Ekuma2014b} or typical
medium dynamical cluster approximation~\cite{Ekuma2014,Ekuma2015}
methods with geometrical average where the density of states
$\rho(\omega=0)$ act as order parameter and disappear at the Fermi
level, such a technique is still useful to investigate the trends
showed by the system in the different regions of parameter space as we
argue below. In the absence of a well defined order parameter in this
approximation, a PM Anderson insulator behavior can be identified in
terms of a gapless flat density of states near the Fermi level.

\begin{figure}
\includegraphics[width=0.98\columnwidth]{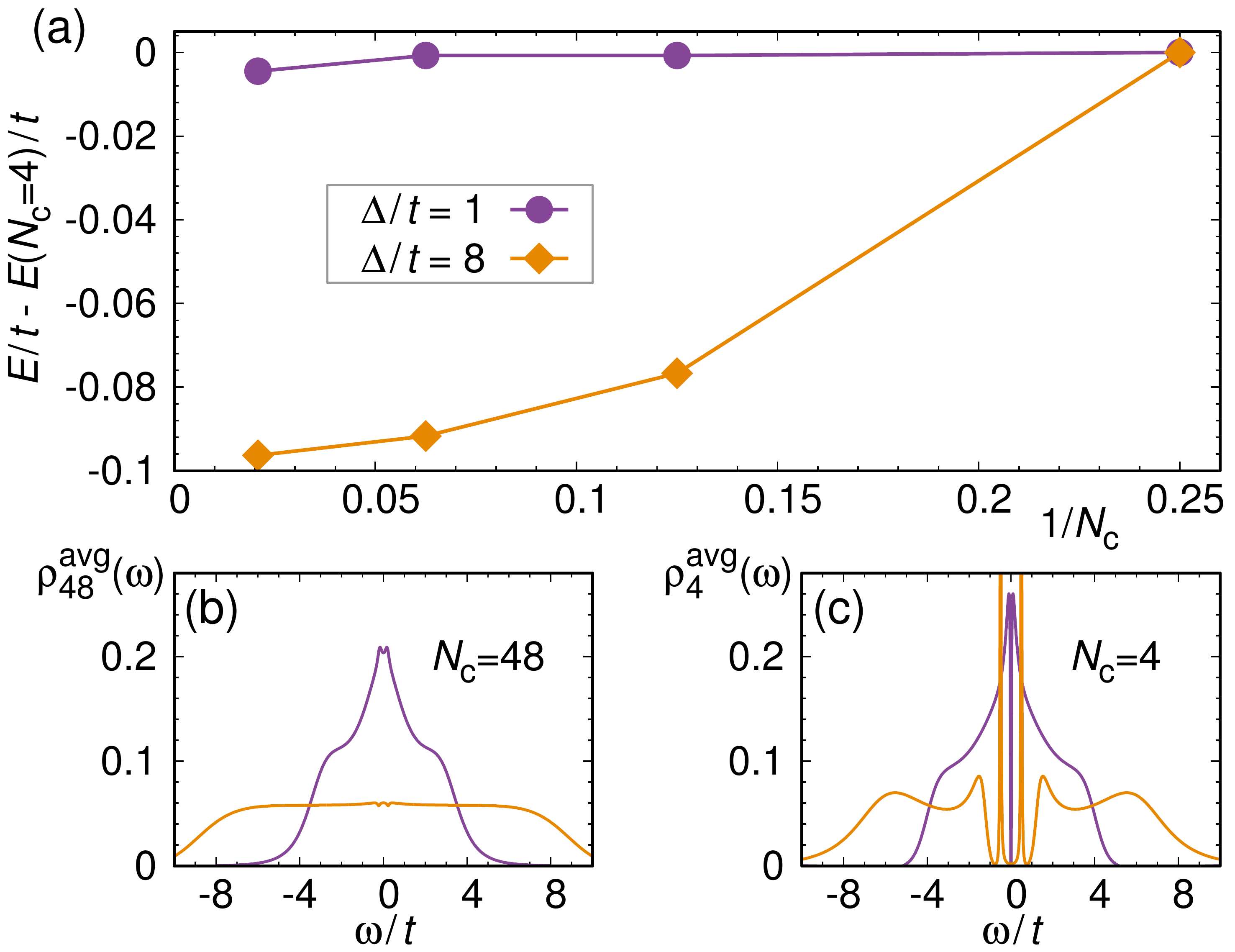}
\caption {\label{Fig1} (Color online) (a) Energies $E/t - E(N_c =
  4)/t$ per site as a function of inverse plaquette size $N_{\rm
    c}^{-1}$ at $\Delta/t=1$ and $8$ for the non-interacting systems,
  where $E(N_c = 4)/t$ are the energies per site at $N_c=4$ for $\Delta/t=1$ and $8$.  The values
  of $E/t - E(N_c = 4)/t$ hardly change with decreasing $N_{\rm
    c}^{-1}$ in the weakly disordered regime $\Delta/t=1$, while they
  continuously decrease in the strongly disordered regime
  $\Delta/t=8$. (b) and (c) show the $\rho^{\text{avg}}_{N_{\rm c}}
  (\omega)$ at $N_{\rm c}=4$ and $48$ for $\Delta/t=1$ and $8$
  obtained from the Pad{\'e} approximation.}
\end{figure}

We first discuss the energies $E$ of the non-interacting disordered
plaquette systems with increasing $N_{\rm c}$ for several $\Delta/t$
obtained from $E=\int d\omega\, \omega \rho_{N_{\rm c}}^{\text{avg}}
(\omega) $ where $\rho_{N_{\rm c}}^{\text{avg}} (\omega)$ is the
averaged density of states obtained within CPA at $N_{\rm
  c}$. Fig.~\ref{Fig1} (a) shows the values of $E/t - E(N_c = 4)/t$ as a
function of $N_{\rm c}^{-1}$ for $\Delta / t = 1$ and $\Delta / t =
8$, where $E(N_c = 4)$ are the energies per site at $N_c=4$ for $\Delta/t=1$ and $8$.  
We observe two different behaviors; (i) at small random disorder ($\Delta / t =
1$)~ $E/t - E(N_c = 4)/t$ remains rather constant as a function of
system size $N_{\rm c}$, while (ii) at large disorder values
$\Delta/t=8$, the energy continuously decreases with increasing
$N_{\rm c}$. Our interpretation is that in the case (i), energy states
around the Fermi level in a pure system would be only slightly
perturbed by impurities and since the Anderson localization length is
expected to be much larger than the plaquette size $N_c$, the energy
values are barely changed with increasing $N_c$.  A PM metallic state
is then realized where the electrons around the Fermi level are easily
pushed into high energy states by small random fluctuations. In this
limit of weak disorder the finite plaquette size impedes observing the
Anderson insulator expected in the non-interacting case in two
dimensions at all disorder strengths.  However, for strong random
disorder ($\Delta / t = 8$) (case (ii)) the band insulator at small
system sizes $N_{\rm c}$ is manifestly different from the appearance
of a PM Anderson insulator at large system sizes.

In order to investigate the differences between cluster sizes $N_{\rm
  c}=48$ and $N_{\rm c}=4$ in more detail, we also plot the
$\rho^{\text{avg}}_{N_{\rm c}}(\omega)$ at $N_{\rm c}=48$ and $4$ for
$\Delta/t=1$ and $8$ in Fig.~\ref{Fig1}~(b) and (c), respectively.  At
$N_{\rm c}=4$, both for $\Delta/t=1$ and $8$ the band insulator is
realized.  At $N_{\rm c}=48$ the slightly perturbed van-Hove
singularity at the Fermi level is still present in the weakly
fluctuating regime ($\Delta/t=1$) as it is the case for $\Delta/t=0$
showing a PM metal behavior, while in the strongly fluctuating regime
$\Delta/t=8$ the van-Hove singularity is completely absent and
$\rho^{\text{avg}} (\omega)$ shows a flat behavior around the Fermi
level, as expected for a PM Anderson insulator~\cite{Shinaoka2009}.

\begin{figure}
\includegraphics[width=1.01\columnwidth]{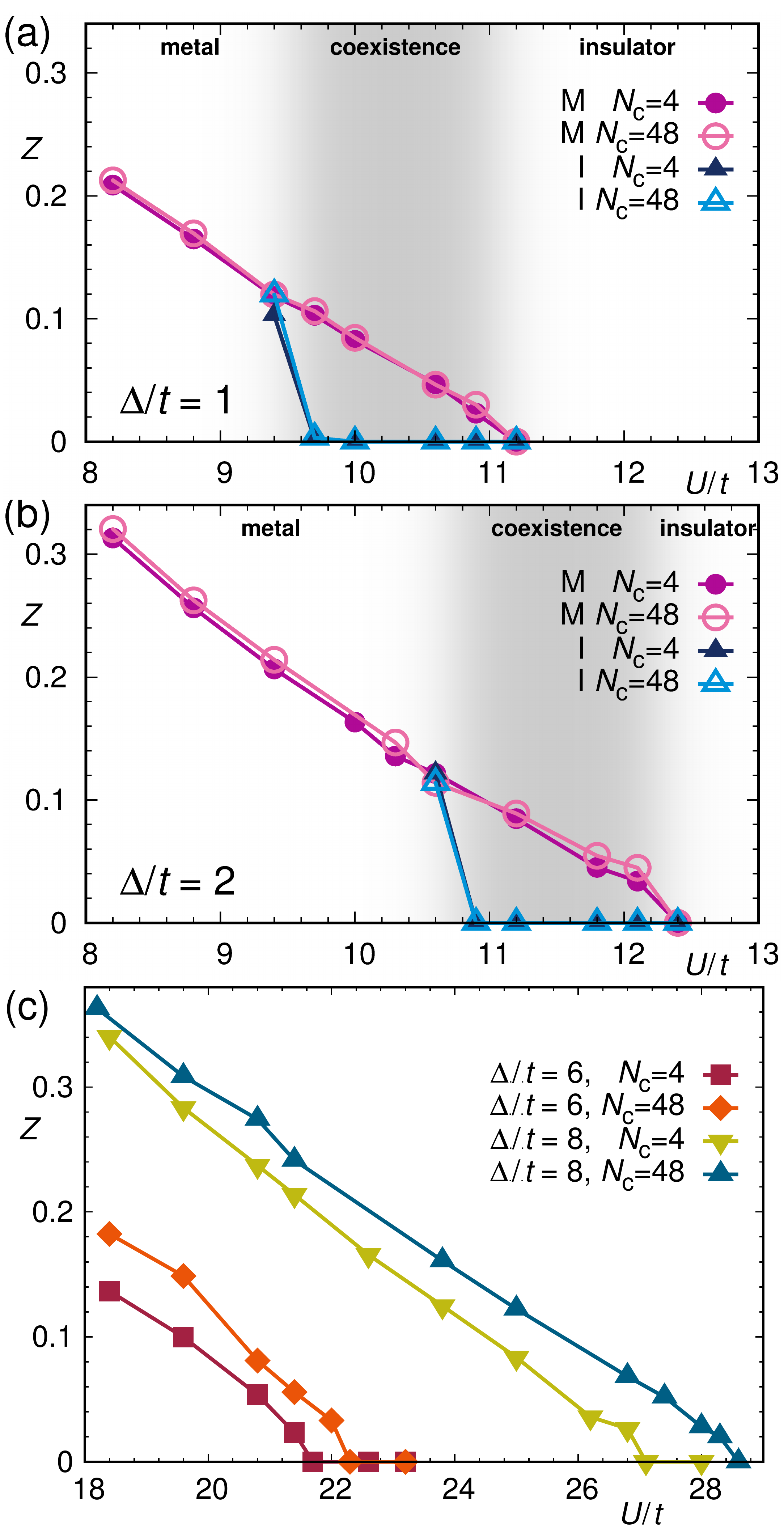}
\caption {\label{Fig2} (Color online) (a) and (b) The quasiparticle
  weight $Z$ as a function of $U/t$ at $N_{\rm c}=4$ and $48$ for
  $\Delta/t=1$ and $2$, respectively.  $T/t$= $0.025$ for the
  single-site dynamical mean field theory (DMFT) calculations.  "M"
  and "I" stand for paramagnetic metal and Mott insulator,
  respectively.  First-order PM metal to Mott insulator transitions,
  where the coexistence regimes including both metal and insulator are
  clearly seen, are present in all cases with small $\Delta/t$. For
  both $N_{\rm c}=4$ and $48$, the critical interactions at
  $\Delta/t=1$ are $U_{c_1}/t = 9.6$ and $U_{c_2}/t = 11.1$; at
  $\Delta/t=2$ they are $U_{c_1}/t = 10.8$ $U_{c_2}/t=12.3$. The
  critical interactions $U_{c_1}$ and $U_{c_2}$ are obtained by
  starting, respectively, from the insulating and metallic solutions
  as initial Weiss field. These critical interaction values are
  comparable to the critical interactions $U_{c_1}/t = 9.4$ and
  $U_{c_2}/t = 10.4$ obtained from single-site DMFT calculations in
  the pure two-dimensional Hubbard model on the square
  lattice~\cite{Park2008}. (c) The quasiparticle weight $Z$ as a
  function of $U/t$ at $N_{\rm c}=4$ and $48$ for $\Delta/t=6$ and
  $8$. The coexistence regions are not observed in all cases with
  large $\Delta/t$.}
\end{figure}

\subsection{Interacting disordered systems}
We now proceed with the disordered interacting systems with increasing
plaquette size $N_{\rm c}$.  We employ the single-site DMFT approach
with temperature $T/t=0.025$ for these
calculations~\cite{Georges1996}.  The DMFT self-consistent equation is
given by
\begin{equation}
G_{N_{\rm c}, \sigma} (i\omega_n) = \int d\epsilon \frac{\rho_{N_{\rm c}}^{\text{avg}} 
(\epsilon)}{i\omega_n + \epsilon - \Sigma_{N_{\rm c}, \sigma} (i\omega_n)},
\end{equation}   
where $\omega_n$ is the Matsubara frequency and the $\rho_{N_{\rm
    c}}^{\text{avg}} (\epsilon)$ is the averaged density of states
over the cluster of size $N_{\rm c} \times N_{\rm c}$ obtained from
the CPA approach.

We consider first the weakly fluctuating disordered regime.
Fig.~\ref{Fig2} (a) and (b) show the quasiparticle weight $Z = ({1 -
  \frac{\rm {Im}(\Sigma(i\omega_0))}{\omega_0}})^{-1}$ as a function
of $U/t$ for systems with two plaquette sizes $N_{\rm c} = 4$ with
ionic potential and $N_{\rm c} = 48$ with random disorder for
$\Delta/t=1$ and $\Delta/t=2$, respectively.  A first-order PM metal
to Mott insulator transition appears in all cases and the coexistence
regimes include both PM metal and Mott insulator behavior. We also
reproduce previous results for the critical interaction
$U_{c_1}/t=9.4$ $(U_{c_2}/t=10.4)$ of former single-site DMFT on the
square lattice for $\Delta/t=0$~\cite{Park2008}.  The critical $U$
values increase with disorder strength to $U_{c_1}/t = 9.6$
$(U_{c_2}/t = 11.1)$ for $\Delta/t= 1$ and $U_{c_1}/t = 10.8~
(U_{c_2}/t=12.3)$ for $\Delta/t=2$ and are almost independent of the
system size with only tiny differences between the $Z$ values for
$N_{\rm c}=4$ and $N_{\rm c}=48$.  These results suggest that if a
weak ionic potential (as the case $N_{\rm c}=4$)
or disorder (as the case $N_{\rm c}=48$) are included in a pure DMFT system
with moderate Coulomb interaction, any effect driven by the ionic
potential or by the random disorder is strongly mitigated by the
strong frustration between the averaged local impurity and the PM bath
in DMFT. Therefore, even though in a non-interacting system a weak
ionic potential ($N_{\rm c}=4$) induces a gapped insulator 
(see Fig.~\ref{Fig1}) and, random
disorder ($N_{\rm c}=48$) induces an Anderson insulator, the physics
exhibited in all weakly interacting
systems with small $\Delta/t$ values would be in
agreement with the single-site DMFT results of a pure system without
random disorder $\Delta/t$. In this weakly interacting and weakly
disordered regime, strong frustration effects between impurity sites
and paramagnetic bath lead to the metallic state.

Next, we discuss the behavior of the electronic correlated system in
the strongly fluctuating disordered regime where a band insulator with
a relatively large gap ($N_{\rm c}=4$) and a PM Anderson insulator
($N_{\rm c}=48$) with a continuous broadened energy band around the
Fermi level in the CPA approach are realized in the non-interacting
case.  We first investigate the behavior of the imaginary part of the
self-energy ${\rm Im}\,\Sigma(i\omega_0)$ and the quasiparticle weight
$Z$.  Fig.~\ref{Fig2} (c) shows $Z$ as a function of $U/t$ for
$\Delta/t=6$ and $\Delta/t=8$. $Z$ decreases with increasing $U/t$ in
all cases.  However for $N_{\rm c}=48$ the Mott insulator behavior
 ($Z$= 0) 
appears at larger $U/t$ values than for $N_{\rm c}=4$. Moreover, we have not observed 
a coexistence regime, unlike the results observed for
the first-order PM metal to Mott insulator transition in the weakly
random disordered regime in Fig.~\ref{Fig2} (a) and (b).

\begin{figure}
\includegraphics[width=0.98\columnwidth]{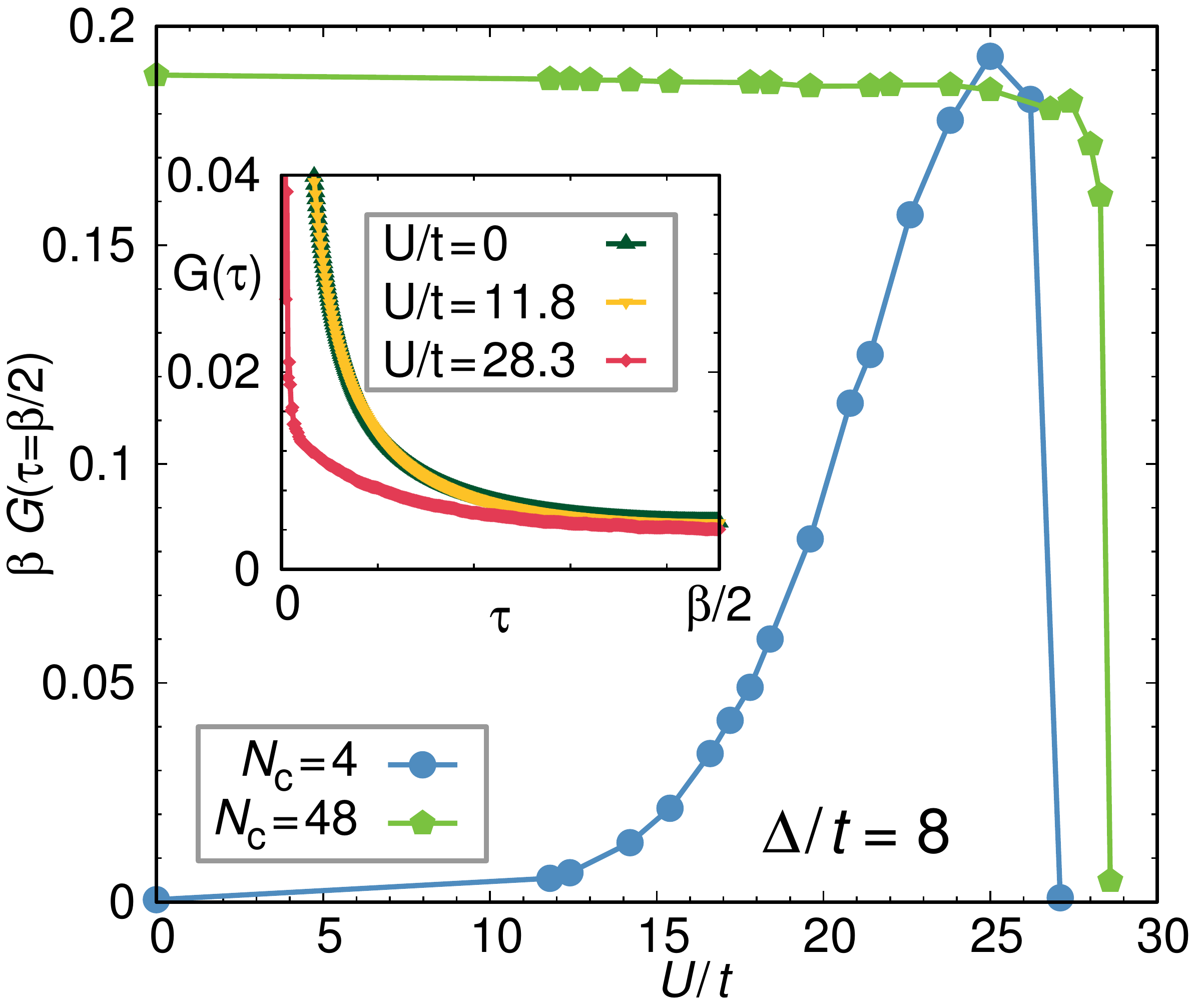}
\caption {\label{Fig3} (Color online) $\beta G(\tau = \beta / 2)$ as a
  function of $U/t$ at $\Delta / t=8$ and $\beta t=40$ for $N_{\rm
    c}=4$ and $48$. $\beta G(\tau = \beta / 2)$ approximates $\rho
  (\omega = 0)$, where $\omega = 0$ is the Fermi level. The inset
  shows $G(\tau)$ for $N_{\rm c}=48$ and $\Delta / t=8$ at three
  different interaction values $U/t=0$, $11.8$ and $28.3$.}
\end{figure}

In the following, since $\beta G(\tau = \beta / 2)$ is approximately
equal to the energy density at the Fermi level $(\beta G(\tau = \beta
/ 2) \approx \rho (\omega = 0))$, we would like to check by computing
$\beta G(\tau = \beta / 2)$, whether the system shows a gapped or
gapless state at the Fermi level ($\omega=0$).  In Fig.~\ref{Fig3}
$\beta G(\tau = \beta / 2)$ is plotted as a function of $U/t$ at
$\Delta /t = 8$ for both system sizes $N_{\rm c}=4$ and $N_{\rm
  c}=48$.  The non-interacting system with $N_{\rm c}=4$ at
$\Delta/t=8$ is a band insulator with an energy gap of $1.0$ and, as
expected, $\beta G(\tau = \beta / 2)$ indicates an insulating behavior
at $U/t=0$.  The gapped state remains up to $U/t=12$, where the
electrons with up and down spin simultaneously occupy the lowest
energy sites.  For $U/t >12$, the electron with spin up (or down)
pushes the electron with spin down (or up) at the same site to 
high energy levels due to the repulsive Coulomb interaction and these
'pushed up' electrons are freely moving in the high energy
levels. This creates a metallic state in the moderately interacting
regions between $U/t=14$ and $27$ by competition and cooperation of
$\Delta/t$ and $U/t$. For large $U/t > 27$, the system becomes a Mott
insulator and $\beta G(\tau = \beta / 2)$ converges to zero.

\begin{figure}
\includegraphics[width=0.98\columnwidth]{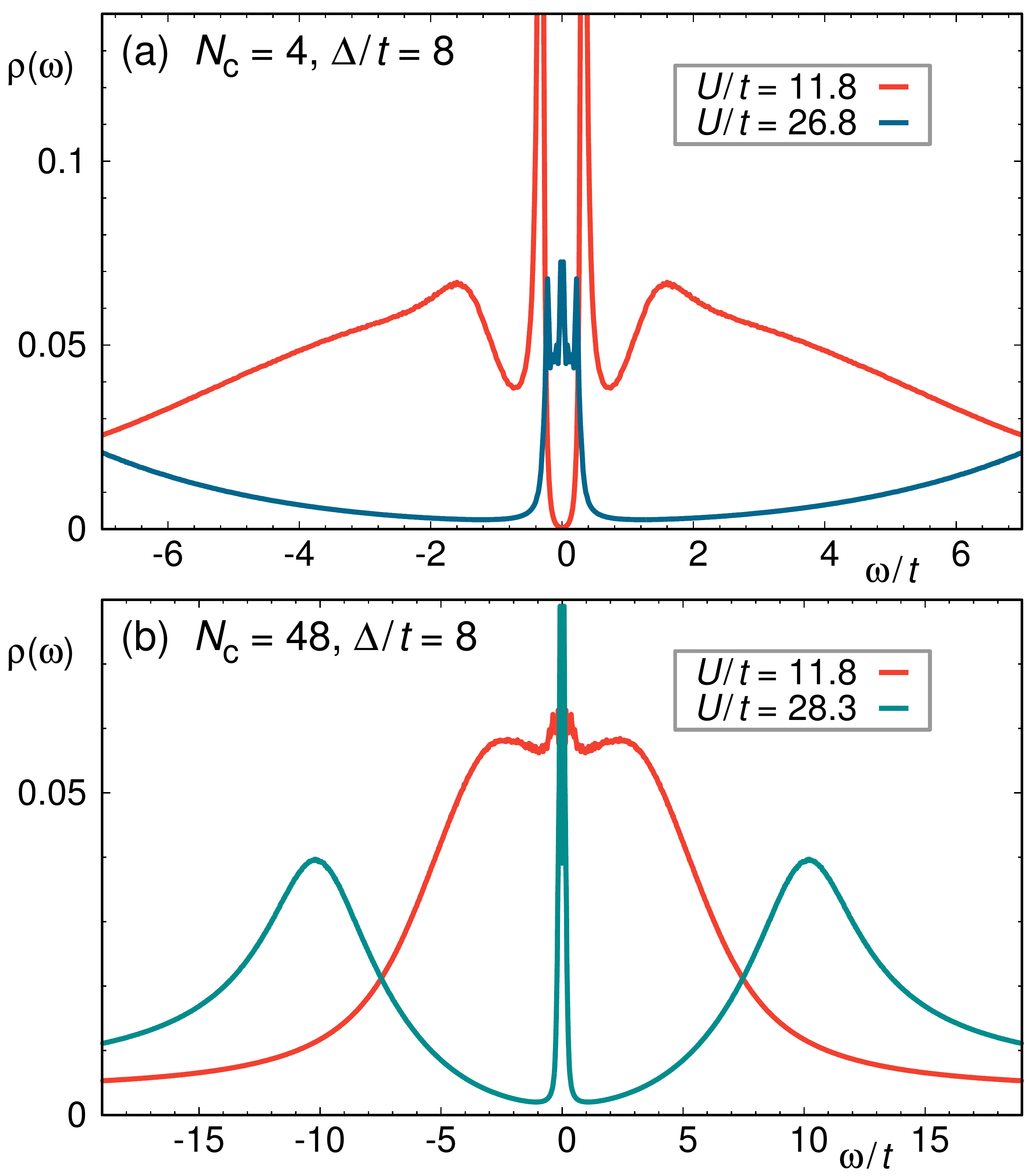}
\caption {\label{Fig4} (Color online) (a) and (b) present the density
  of states $\rho(\omega)$ of systems with $\Delta/t = 8$, $U/t=11.8$
  and $26.8$ for $N_{\rm c}=4$ and $U/t=11.8$ and $28.3$ for $N_{\rm
    c}=48$ under the condition of $\rho(\omega) = \rho(-\omega)$ by
  the particle-hole symmetry, respectively.}
\end{figure}

The case of $N_{\rm c}=48$ is distinctly different.  At the value
$\Delta /t = 8$ a PM Anderson insulator may be expected for the
non-interacting system. As $U/t$ increases, the value $\beta G(\tau =
\beta / 2)= 0.187$ remains unchanged up to $U/t=25$. This behavior
suggests that the physical state corresponding to a PM Anderson
insulator remains up to this $U/t$ value. In the inset of
Fig.~\ref{Fig3} we have plotted $G(\tau)$ for $U/t=0$, $11.8$, and
$28.3$ in order to confirm this suspicion.  We confirm that $G(\tau)$
for $U/t=0$ and $11.8$ fall on top of each other.  This may indicate
that for strong disorder the PM Anderson insulator is still preserved
in the weakly interacting regime, even though interactions are
involved.  Around $U/t=28.3$, $\beta G(\tau = \beta / 2)$ shows a kink
that hints to the presence of a correlated metallic state due to the
fact that interactions push up the electrons of low states into high
states. In the strongly correlated regime $U/t > 28.3$ in
Fig.~\ref{Fig3}, $\beta G(\tau = \beta / 2)$ converges to zero which
indicates the Anderson-Mott insulator behavior.

\begin{figure}
\includegraphics[width=0.98\columnwidth]{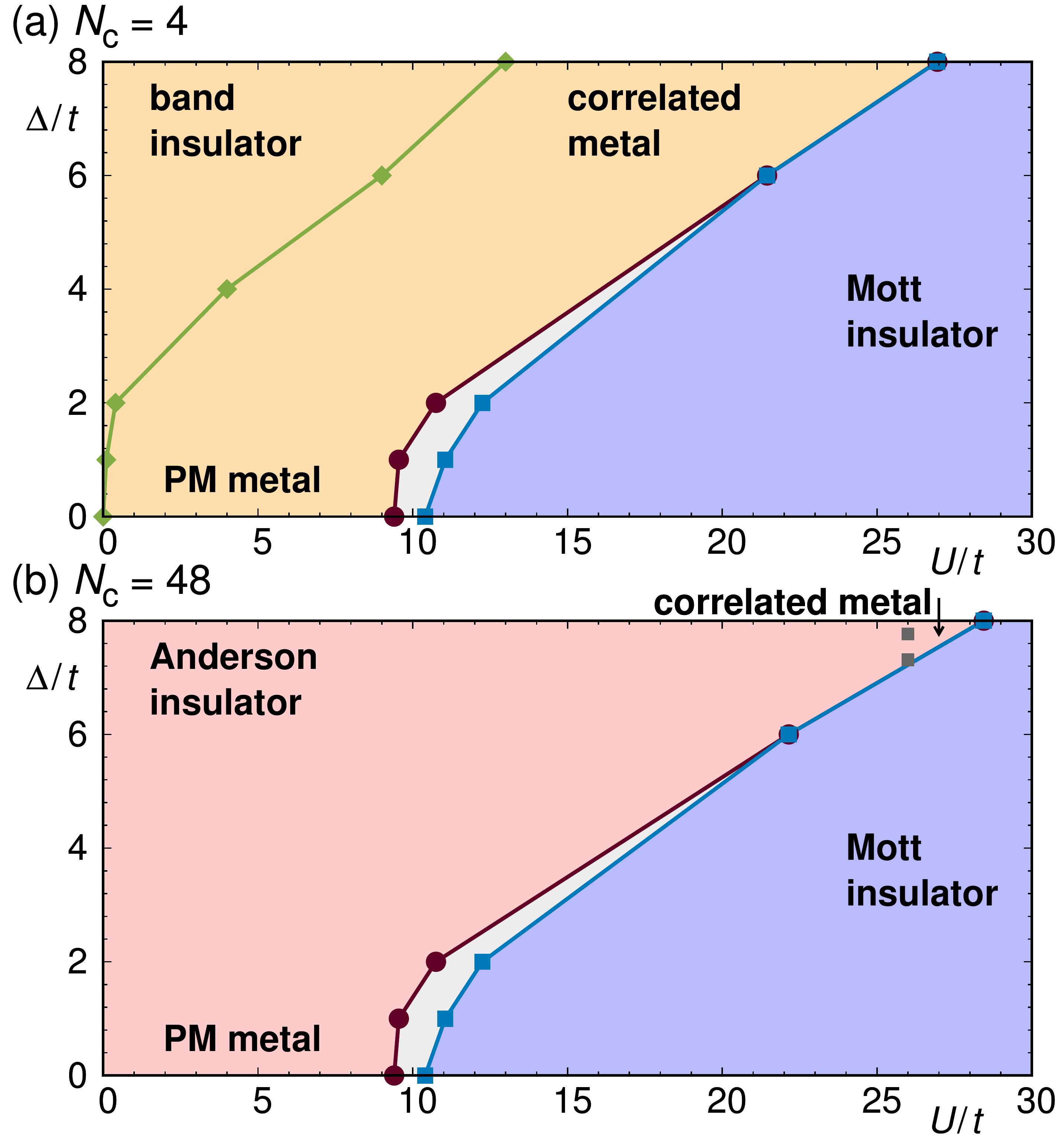}
\caption {\label{Fig5} (Color online) Phase diagrams of the
  two-dimensional $N_{\rm c} \times N_{\rm c}$ plaquette disordered
  interacting systems at (a) $N_c=4$ and (b) $48$ via the single-site
  dynamical mean field theory. The coexistence regions including
  behaviors of both PM metal and Mott insulator are located between
  the PM metal and the Mott insulator in the weakly disordered
  regions. The critical interaction $U_c$ is obtained by sweeping in
  steps of 0.4 the value of the interaction $U$. The critical
  interactions $U_c$ at $\Delta/t=0$ are obtained from
  Ref.~\protect\onlinecite{Park2008}.  The error bars are given by the
  size of the symbols. Note that the determination of the crossover
  line between PM metal and Anderson insulator for the case of $N_c=
  48$ in (b) is much more difficult since the physical quantities that
  account for an accurate phase boundary between Anderson insulator
  and PM metal are absent within CPA. Therefore, we determine the
  correlated metallic regimes, where the electrons in low energy
  states push each other into high energy states, and phase boundary
  only in the strong disorder region with $\Delta/t=8$ and $N_c=48$
  in (b), via estimation of $\beta G(\tau=\beta/2)$ of
  Fig.~\ref{Fig3}.}
\end{figure}

In order to establish more clearly the various physical phases, we
explore $\rho (\omega)$ calculated by a stochastic analytical
continuation from $G(i\omega_n)$~\cite{Beach2004}.  Fig.~\ref{Fig4} (a) and (b) show
$\rho (\omega)$ at $\Delta/t=8$ and various $U/t$ for $N_{\rm c}=4$
and $N_{\rm c}=48$, respectively. $\rho (\omega)$ for $N_{\rm c}=4$
(Fig.~\ref{Fig4} (a)) exhibits a band insulator at $U/t=11.8$ and a
Fermi liquid behavior with a quasiparticle peak at $U/t=26.8$ and
agrees with the results of the density of states at the Fermi level
estimate $\beta G(\tau=\beta / 2)$ in Fig.~\ref{Fig3}.  $\rho
(\omega)$ for $N_{\rm c}=48$ (Fig.~\ref{Fig4} (b)) shows flat behavior
around the Fermi level at $U/t=11.8$ and this result is similar to
that at $U/t=0$ (Fig.~\ref{Fig1} (b)) which hints to a PM
Anderson insulator.  At $U/t=28.3$, $\rho (\omega)$ shows a
quasiparticle peak at the Fermi level and a Hubbard band around $\pm
\omega/t=11$.  The repulsive Coulomb interaction pushes the electrons
at low energies into high energy states inducing an Anderson insulator
to correlated metal transition. This region is however very narrow.

\section{Conclusions}

We summarize the results of this study by plotting the phase diagrams
for disordered interacting systems at $N_c=4$ and $48$ in
Fig.~\ref{Fig5} (a) and (b), respectively. Even though the
single-site DMFT approach emphasizes local fluctuations in a
fully frustrated PM bath and thus overestimates the PM metal regions in both
cases, it still captures basic properties of the systems considered
such as band insulator, Anderson insulator, correlated metal, and Mott
insulator induced by random disorder and electronic correlations.  In
more detail, the metal to Mott insulator transition in the weak
disordered regions are closer to those of DMFT results without
disorder because the strong local fluctuations of the DMFT
approximation, most probably, overwhelm effects driven by disorder,
while we find sandwiched {\it correlated metallic states} between a
band insulator and a Mott insulator with $N_c=4$ and between an
Anderson insulator and a Mott insulator with $N_c=48$ at strong
disordered regimes. Such states can be understood by the fact that
electrons in the low energy states push each other into high energy
states and correlated metallic states with a quasiparticle peak and
Hubbard bands emerge.
 
Even though we considered a simple model, the emergence of the various
competing phases as a function of cluster size, electron-electron
interaction and disorder may provide further hints, for instance, to
the microscopic origin of the metal-insulator transition observed in
two-dimensional metal-oxide-semiconductor field-effect
transistors~\cite{Kravchenko1996,Lin2015}. Furthermore, confinement of Fermionic
atoms in an optical lattice to realize ionic-type Hubbard models is
conceivable and could be guided by our results for small $N_c$.

Finally, in view of the  physics uncovered in this work
for the two-dimensional plaquette Hubbard model, 
a next step would be, when very fast multi-site impurity
solvers become available,  to employ the more advanced 
 recently developed typical medium dynamical cluster
approximation~\cite{Ekuma2014,Ekuma2014b} which can account for nonlocal spatial
correlations and for the Anderson localization length beyond the CPA in combination with the DMFT
approach.

\begin{acknowledgments}
H.L. is supported by the Korea-German researcher exchange program
(NRF-2014K2A5A5030523).  H.O.J. and R.V. gratefully acknowledge financial
support from the Deutsche Forschungsgemeinschaft through grant FOR~1346.
\end{acknowledgments}


\begin{thebibliography}{99}

\bibitem{Imada1998} M. Imada, A. Fujimori, Y. Tokura, Metal-insulator
  transitions, Rev. Mod. Phys. {\bf 70}, 1039 (1998).

\bibitem{Evers2008} F. Evers, A. D. Mirlin, Anderson transitions,
  Rev. Mod. Phys. {\bf 80}, 1355 (2008).

\bibitem{Anderson1958} P. W. Anderson, Absence of Diffusion in Certain
  Random Lattices, Phys. Rev. {\bf 109}, 1492 (1958).

\bibitem{Abrahams1979} E. Abrahams, P. W. Anderson,
  D. C. Licciardello, T. V. Ramakrishnan, Scaling Theory of
  Localization: Absence of Quantum Diffusion in Two Dimensions,
  Phys. Rev. Lett. {\bf 42}, 673 (1979).

\bibitem{Kramer1993} B. Kramer, A. MacKinnon, Localization: theory and
  experiment, Rep. Prog. Phys. {\bf 56}, 1469 (1993).

\bibitem{Slevin1999} K. Slevin, T. Ohtsuki, Corrections to Scaling at
  the Anderson Transition, Phys. Rev. Lett. {\bf 82}, 382 (1999).

\bibitem{Ekuma2014} C. E. Ekuma, H. Terleska, K.-M. Tam, Z.-Y Meng,
  J. Mereno, M. Jarrell, Typical medium dynamical cluster
  approximation for the study of Anderson localization in three
  dimensions, Phys. Rev. B {\bf 89}, 081107(R) (2014).

\bibitem{Ekuma2014b} C. E. Ekuma, H. Terletska, Z. Y. Meng, J. Moreno,
  M. Jarrell, S. Mahmoudian and V. Dobrosavljevi{\'c},
  J. Phys. Condens Matter {\bf 26}, 274209 (2014).

\bibitem{Dobrosavljevic1997} V. Dobrosavljevic, G. Kotliar, Mean Field
  Theory of the Mott-Anderson Transition, Phys. Rev. Lett. {\bf 78},
  3943 (1997).

\bibitem{Byczuk2005} K. Byczuk, W. Hofstetter, D. Vollhardt,
  Mott-Hubbard Transition versus Anderson Localization in Correlated
  Electron Systems with Disorder, Phys. Rev. Lett. {\bf 94}, 056404
  (2005).

\bibitem{Byczuk2009} K. Byczuk, W. Hofstetter, D. Vollhardt,
  Competition between Anderson Localization and Antiferromagnetism in
  Correlated Lattice Fermion Systems with Disorder,
  Phys. Rev. Lett. {\bf 102}, 146403 (2009).

\bibitem{Heidarian2004} D. Heidarian, N. Trivedi, Inhomogeneous
  Metallic Phase in a Disordered Mott Insulator in Two Dimensions,
  Phys. Rev. Lett. {\bf 93}, 126401 (2004).

\bibitem{Shinaoka2009} H. Shinaoka, M. Imada, Soft Hubbard Gaps in
  Disordered Itinerant Models with Short-Range Interaction,
  Phys. Rev. Lett. {\bf 102}, 016404 (2009).

\bibitem{Chiesa2008} S. Chiesa, P. B. Chakraborty, W. E. Pickett,
  R. T. Scalettar, Disorder-Induced Stabilization of the Pseudogap in
  Strongly Correlated Systems, Phys. Rev. Lett. {\bf 101}, 086401
  (2008).

\bibitem{Oliveira2014} W. S. Oliveira, M. C. O. Aguiar,
  V. Dobrosavljevic, Mott-Anderson transition in disordered
  charge-transfer model: Insights from typical medium theory,
  Phys. Rev. B {\bf 89}, 165138 (2014).

\bibitem{Ekuma2015} C. E. Ekuma, C. Moore, H. Terletska, K.-M. Tam,
  J. Moreno, M. Jarrell, N. S. Vidhyadhiraja, Finite-cluster typical
  medium theory for disordered electronic systems, Phys. Rev. B {\bf
    92}, 014209 (2015).

\bibitem{Dobrosavljevic2003} V. Dobrosavljevic, A. A. Pastor,
  B. K. Nikolic, Typical medium theory of Anderson localization: A
  local order parameter approach to strong-disorder effects,
  Europhys. Lett. {\bf 62}, 76 (2003).

\bibitem{Kim2005} K. W. Kim, J. S. Lee, T. W. Noh, S. R. Lee, K. Char,
  Metal-insulator transition in a disordered and correlated
  SrTi$_{1-x}$Ru$_x$O$_3$ system: Changes in transport properties,
  optical spectra, and electronic structure, Phys. Rev. B {\bf 71},
  125104 (2005).

\bibitem{Maiti2007} K. Maiti, R. S. Singh, V. R. R. Medicherla,
  Evolution of a band insulating phase from a correlated metallic
  phase, Phys. Rev. B {\bf 76}, 165128 (2007).

\bibitem{Lahoud2014} E. Lahoud, O. Nganba Meetei, K. B. Chaska,
  N. Trivedi, Emergence of a Novel Pseudogap Metallic State in a
  Disordered 2D Mott Insulator, Phys. Rev. Lett. {\bf 112}, 206402
  (2014).

\bibitem{Chikara2015} S, Chikara, D. Haskel, J.-H. Kim, H.-S Kim, C.-C
  Chen, G. Fabbris, L. S. V. Viega, N. M. Souza-Neto, J. Terzic,
  K. Butrouna, G. Cao, M. J. Han, M. van Veenendaal,
  Sr$_2$Ir$_{1-x}$Rh$_x$O$_4$($x<0.5$): An inhomogeneous $j_{\rm
    eff}=\frac{1}{2}$ Hubbard system, Phys. Rev. B {\bf 92}, 081114(R)
  (2015).

\bibitem{Muthu2012} K. Muthukumar, H. O. Jeschke, R. Valent\'\i,
  E. Begun, J. Schwenk, F. Porrati, M. Huth, Spontaneous Dissociation
  of Co$_2$(CO)$_8$ and Autocatalytic growth of Co on SiO$_2$ : A
  Combined Experimental and Theoretical Investigation, Beilstein
  J. Nanotech. {\bf 3}, 546 (2012).

\bibitem{Muthu2012_2} K. Muthukumar, R. Valent\'\i, H. O. Jeschke,
  Simulation of structural and electronic properties of amorphous
  tungsten oxycarbides, New. J. Phys. {\bf 14}, 113028 (2012).

\bibitem{Kravchenko1996} S. V. Kravchenko, D. Simonian,
  M. P. Sarachik, W. Mason, J. E. Furneaux, Electric Field Scaling at
  a $B=0$ Metal-Insulator Transition in Two Dimensions,
  Phys. Rev. Lett. {\bf 77}, 4938 (1996).

\bibitem{Lin2015} Ping V. Lin and Dragana Popovic, Critical Behavior
  of a Strongly Disordered 2D Electron System: The Cases of Long-Range
  and Screened Coulomb Interactions, Phys. Rev. Lett. {\bf 114},
  166401 (2015).

\bibitem{Kancharla2007} S. S. Kancharla, E. Dagotto, Correlated
  Insulated Phase Suggests Bond Order between Band and Mott Insulators
  in Two Dimensions, Phys. Rev. Lett. {\bf 98}, 016402 (2007).

\bibitem{Paris2007} N. Paris, K. Bouadim, F. Hebert, G.G. Batrouni,
  R. T. Scalettar, Quantum Monte Carlo Study of an Interaction-Driven
  Band-Insulator-to-Metal Transition, Phys. Rev. Lett. {\bf 98},
  046403 (2007).

\bibitem{Go2011} A. Go, G. S. Jeon, Phase transitions and spectral
  properties of the ionic Hubbard model in one dimension, Phys. Rev. B
  {\bf 84}, 195102 (2011).

\bibitem{Tugushev1996} V. Tugushev, S. Caprara, M. Avignon, Spin-density-wave transition in systems with chemical dimerization, 
Phys. Rev. B {\bf 54}, 5466 (1996).

\bibitem{Caprara2000} S. Caprara, M. Avignon, O. Navarro, Spin-density-wave transition in systems with chemical dimerization, 
Phys. Rev. B {\bf 61}, 15667 (2000).


\bibitem{Georges1996} A. Georges, G. Kotliar, W. Krauth,
  M. J. Rozenberg, Dynamical mean-field theory of strongly correlated
  fermion systems and the limit of infinite dimensions,
  Rev. Mod. Phys. {\bf 68}, 13 (1996).

\bibitem{Park2008} H. Park, K. Haule, G. Kotliar, Cluster Dynamical
  Mean Field Theory of the Mott Transition, Phys. Rev. Lett. {\bf
    101}, 186403 (2008).

\bibitem{Rubtsov2005} A. N. Rubtsov, V. V. Savkin, A. I. Lichtenstein,
  Continuous-time quantum Monte Carlo method for fermions,
  Phys. Rev. B {\bf 72}, 035122 (2005).

\bibitem{Gull2011} E. Gull, A. J. Millis, A. I. Lichtenstein,
  A. N. Rubtsov, M. Troyer, P. Werner, Continuous-time Monte Carlo
  methods for quantum impurity models, Rev. Mod. Phys. {\bf 83}, 349
  (2011).

\bibitem{Lee2014} H. Lee, Y.-Z. Zhang, H. O. Jeschke, R. Valent\'\i,
  Competition between band and Mott insulator in the bilayer Hubbard
  model: a dynamical cluster approximation study, Phys. Rev. B {\bf
    89}, 035139 (2014).

\bibitem{Soven1967} P. Soven, Coherent-Potential Model of
  Substitutional Disordered Alloys, Phys. Rev. {\bf 156}, 809 (1967).

\bibitem{Beach2004} K. S. D. Beach, Identifying the maximum entropy method as a special limit of stochastic analytic continuation, 
arXiv:cond-mat/0403055 (2004).



\end{thebibliography}
\end{document}